\documentclass[useAMS,usenatbib]{mn2e}
\usepackage{graphicx}
\usepackage{longtable}
\usepackage{amssymb}

\title[Gas content in AGN and galaxies]{Black hole accretion preferentially occurs in gas rich galaxies\thanks{{\it Herschel} is an ESA space observatory with science instruments provided by European-led Principal Investigator consortia and with important participation from NASA.}}
\author[F. Vito et al.]
{F. Vito$^{1,2,3,4}$\thanks{E-mail: fabio.vito@unibo.it },
R. Maiolino$^{3,4}$,
P. Santini$^{5}$,
M. Brusa$^{1,2,6}$,
A. Comastri$^{2}$,
G. Cresci$^{7}$,
\newauthor
D. Farrah$^8$,
A. Franceschini$^9$,
R. Gilli$^2$,
G. L. Granato$^{10}$, 
C. Gruppioni$^2$, 
D. Lutz$^6$,
\newauthor
F. Mannucci$^7$,
F. Pozzi$^1$,
D. J. Rosario$^6$,
D. Scott$^{11}$,
M. Viero$^{12}$,
C. Vignali$^1$
\\ \\
$^1$ Dipartimento di Fisica e Astronomia, Universit\`a degli Studi di Bologna, Via Ranzani 1, 40127 Bologna, Italy \\
$^2$ INAF -- Osservatorio Astronomico di Bologna, Via Ranzani 1, 40127 Bologna, Italy\\
$^3$ Cavendish Laboratory, University of Cambridge, 19 J. J. Thomson Avenue, Cambridge CB3 0HE, UK\\
$^4$ Kavli Institute for Cosmology, University of Cambridge, Madingley Road, Cambridge CB3 0HA, UK \\
$^5$ INAF -- Osservatorio Astronomico di Roma, via di Frascati 33, 00040 Monte Porzio Catone, Italy\\
$^6$ Max-Planck-Institut f\"{u}r extraterrestrische Physik (MPE), Giessenbachstrasse 1, D-85748, Garching bei M\"{u}nchen, Germany\\
$^7$ INAF -- Osservatorio Astrofisico di Arcetri, Largo E. Fermi 5, 50125 Firenze, Italy\\
$^8$ Department of Physics, Virginia Tech, Blacksburg, VA 24061, USA\\
$^9$ Dipartimento di Fisica e Astronomia, Universit\`a di Padova, vicolo Osservatorio, 3, I-35122 Padova, Italy\\
$^{10}$ INAF -- Osservatorio Astronomico di Trieste, via Tiepolo 11, I-34131 Trieste, Italy\\
$^{11}$ Department of Physics and Astronomy, University of British Columbia, 6224 Agricultural Road, Vancouver, BC V6T 1Z1, Canada\\
$^{12}$ California Institute of Technology, 1200 E. California Blvd., Pasadena, CA 91125, USA
}
\begin{document}

\date{Submitted 23 January 2014. Accepted 31 March 2014.}

\graphicspath{{.}}
% \date{}

\pagerange{\pageref{firstpage}--\pageref{lastpage}} \pubyear{2014}

\maketitle

\label{firstpage}

\begin{abstract}
We have investigated the gas content of a sample of several hundred AGN host galaxies
at z$<$1 and compared it with a sample of inactive galaxies, matched
in bins of stellar mass and redshift.
Gas masses have been inferred from the dust masses, obtained by stacked Herschel far-IR and sub-mm
data in the GOODS and COSMOS fields, under reasonable assumptions and metallicity
scaling relations for the dust-to-gas ratio. We find that AGNs are on average hosted in galaxies much more
gas rich than inactive galaxies. In the vast majority of stellar mass bins, the average gas content of AGN hosts is
higher than in inactive galaxies. The difference is up to a factor of ten
higher in low stellar mass galaxies, with a significance of 6.5$\sigma$. In almost half of the AGN sample the
gas content is three times higher than in the control sample of inactive galaxies.
Our result strongly suggests that the probability of having an AGN activated is simply driven
by the amount of gas in the host galaxy; this can be explained
in simple terms of statistical probability of 
having a gas cloud falling into the gravitational potential of the black hole.
The increased probability of an AGN being hosted by a star-forming galaxy, identified by previous works, may be a
consequence of the relationship between gas content and AGN activity, found
in this paper, combined with the Schmidt-Kennicutt law for star formation.
\end{abstract}


\section{Introduction}\label{sec1}
During the past decades
several studies have revealed
a connection between the mass of Black Holes hosted in galactic nuclei
and the properties of their host galaxies
\citep[e.g.][]{Magorrian98, Ferrarese00, Marconi03}. Moreover, the redshift evolution of the
cosmic Star Formation Rate (SFR) and SMBH accretion rate density are very similar
\citep{Boyle98,Granato01,Marconi04,Hopkins06,Silverman09,Aird10}.
This ``co-evolution" has led various authors to investigate a possible connection between the presence of Active Galactic Nuclei (AGN, which traces the SMBH growth) and the star formation properties of galaxies.
A correlation was indeed observed between star formation and nuclear activity in numerous works at
high AGN luminosities \citep[e.g][]{Lutz08, Lutz10, Shao10, Rosario12}, while at low AGN
luminosities this link is more debated, with \cite{Silverman09} reporting no significant difference
in the SFR between active and inactive galaxies, while \cite{Santini12} found a slight enhancement
for AGN-hosting galaxies. They also reported that the enhancement of star formation activity in AGN
with respect to the bulk of inactive galaxies disappeared if quiescent galaxies were discarded, i.e. AGN are more likely hosted in star forming galaxies.

The gas content is often regarded as a
more fundamental property of galaxies, with respect to the SFR.
The SFR is tightly related to the gas content through the Schmidt-Kennicutt relation
\citep[SK relation hereafter]{Schmidt59,Kennicutt98}. Currently, one of the most favored scenarios is that the cosmic evolution
of the star formation rate in galaxies is mostly a consequence, through the SK relation,
of the more fundamental evolution of their (molecular) gas
content \citep[e.g.][]{Obreschkow09,Lagos11}.

Within the context of AGNs,
gas is the fundamental ingredient both for nuclear activity and star formation. Possible
differences in terms of star formation
properties between AGNs hosts and inactive galaxies could be due to more fundamental differences in
terms of gas content, as recently argued by \cite{Santini12} and \cite{Rosario12,Rosario13_2}. Therefore, it is most important
to obtain information on the gas content of AGN host galaxies, possibly as a function of galaxy properties (e.g.
stellar mass) and redshift.

The molecular gas content can be inferred from the luminosity of the CO millimeter transitions, by assuming a proper
CO-to-H$_2$ conversion factor. 
However, CO observations are very time consuming, and surveys of large samples
are extremely difficult and time demanding. Alternatively, the total (molecular and atomic) gas mass can be derived from the
dust content, inferred from the FIR-submm SED, by
assuming a dust-to-gas ratio \citep[DGR; e.g.][]{Eales10, Leroy11,Magdis11}. The uncertainties on the
dust-to-gas ratio and its dependence on metallicity are similar to those affecting the CO-to-H$_2$ conversion factor,
making the two methods comparable in terms of accuracy, at least at metallicities 12+log(O/H)$>$8.0
\citep{Bolatto13, Remy-Ruyer13}.

In this work we exploit the dust method for measuring the gas masses in AGN host galaxies. In particular,
gas masses are obtained from the dust mass derived from the FIR SED
of several hundred AGN host galaxies at z$<$1, along with a control sample of normal galaxies
selected in the same stellar mass and redshift ranges.
The aim of this work is to investigate differences in terms of gas content
between AGN hosts and the bulk of the galaxy population (i.e. star forming and quiescent galaxies),
in bins of stellar mass and redshift, to avoid potential biases caused by the dependency of the gas content on these two quantities.
We make use of a stacking procedure to increase the luminosity completeness of the studied samples (\S~\ref{stack}).

\section{Data set and sample selection }\label{sec2}
We selected a sample of AGN at $z\leq 1$ (as well as normal galaxies for the control sample) in the
COSMOS, GOODS-S and GOODS-N fields. The choice of these fields was driven by the wide
multiwavelength coverage provided by a number of surveys, which is crucial to derive reliable
properties (redshift, stellar mass, star-formation rate and dust mass). 

We used the Far-Infrared (FIR) data from the PACS Evolutionary Probe (PEP, \citealt{Lutz11}) and the Herschel Multi-tiered Extra-galactic Survey (HerMES, \citealt{Oliver12}) programs, which cover the three fields used in this work. The former 
was performed with the PACS camera (70, 100 and $160\rmn{\,\mu m}$; \citealt{Poglitsch10}) while the latter with the SPIRE
camera (250, 350 and $500\,\rmn{\mu m}$; \citealt{Griffin10}), both on board of the Herschel Space Observatory
\citep{Pilbratt10}. Herschel catalogues are based on prior information on MIPS $24\,\rmn{\mu m}$ positions and fluxes. The
PEP catalogue was described by \cite{Lutz11} and \cite{Berta11}, while the HerMES catalogue was presented by
\cite{Roseboom10,Roseboom12}. Since only the GOODS-S field was observed at 70 $\rmn{\mu m}$, following \citet*[S13 hereafter]{Santini13}, who reported that the exclusion of that band does not significantly affect the results, we will not use the 70
$\rmn{\mu m}$ data, to use a consistent procedure among all the fields.

For the GOODS-S field we used the optical/near-IR photometric data from the GOODS-MUSIC catalogue
\citep{Grazian06,Santini09} and the X-ray counterpart information from the 4 Ms \textit{Chandra}
Deep Field South (CDF-S) main catalogue \citep{Xue11}. In the GOODS-N field we collected the
multiwavelength data from the PEP team catalogue \citep{Berta11} and the X-ray data from the 2 Ms
\textit{Chandra} Deep Field North catalogue \citep{Alexander03, Bauer04}. Finally, in the COSMOS
field we used the \cite{Ilbert09} and \cite{McCracken10} multiwavelength catalogues and the
\textit{Chandra} \citep{Civano12} and XMM \citep{Brusa10} COSMOS optical identification catalogues,
complemented by the photometric redshifts presented by \cite{Salvato11}. All the catalogues are
supplemented with spectroscopic or photometric redshifts. Photometric redshift for inactive galaxies
in COSMOS and GOODS-N lacking redshift information in the above-mentioned catalogues were computed
by using the EAZY code \citep{Brammer08}.

We applied two selection criteria to the parent sample in the three fields: 1) signal-to-noise ratio (SNR) $\geq 10$ in the
K-band; 2) $z\leq1$. The K-band selection ensures that we can derive reliable estimates for the stellar mass
(following \citeauthor{Santini13}), while the redshift cut ensures that the Herschel bands are not probing rest-frame
wavelengths that may be significantly affected by AGN heating and reprocessing (which may
affect our estimation of the dust masses). Indeed,
\cite{Rosario12} have shown that out to z$\sim$1 the PACS colors are consistent with those
typical of star forming galaxies, while some AGN contamination to the 100$\mu$m may occur at higher
redshift.

Once these selections are applied, we divided the resulting objects into AGN and a galaxy samples. In COSMOS, we considered an X-ray detected object to be an AGN following its optical classification \citep{Brusa10,Civano12} or best-fitting SED template \citep{Salvato11} or if its absorption-corrected luminosity in the rest-frame $2-10\,\rmn{keV}$ band, where available, is $L_X\geq10^{42}\rmn{erg\,s^{-1}}$. This threshold is a compromise between the sample size and the contamination by purely powerful star-forming galaxies \citep{Ranalli03}. The luminosities were collected from \cite{Lanzuisi13} and \cite{Mainieri07, Mainieri11}, who performed a spectral analysis on a sample of bright sources in the \textit{Chandra}-COSMOS \citep{Elvis09} and XMM-COSMOS \citep{Cappelluti09} catalogues, respectively. Intrinsic luminosities of all the XMM-COSMOS selected sources were also derived by the XMM-COSMOS team using the observed fluxes and Hardness Ratio HR \citep[see][]{Merloni13}. In GOODS-S we 
assumed the \cite{Xue11} classification and luminosity for the X-ray sources. 
In GOODS-N we used the X-ray detected AGN sample of \cite{Bauer04}. The luminosities of the X-ray selected AGN for which this information could be retrieved are in the range $10^{41}\lesssim L_X <10^{45}\rmn{erg\,s^{-1}}$, and only $\sim 5$ per cent of them have $L_X >10^{44}\rmn{erg\,s^{-1}}$. Given the limiting fluxes of the X-ray samples, the most obscured AGN can still remain undetected. In order to at least partially recover them, we applied the IRAC power-law selection by \cite{Donley12}. The resulting AGN sample consists in 801 objects, of which 631 X-ray selected ( in any X-ray band considered in the above-mentioned catalogues; 486 in COSMOS and 145 in GOODS-S+N) and
255 IRAC selected (241 in COSMOS and 14 in GOODS-S+N). Eighty-five AGN (78 in COSMOS and 7 in GOODS-S+N) are in common between the two selections. All the sources
which do not satisfy any of the AGN selection criterion in each field are automatically classified as
inactive galaxies ($\sim175000$ objects).

\section{Parameters derivation}\label{parameters}
In this section we describe how the physical parameters for our sample of AGN and galaxies were derived. In order to avoid
selection effects due to redshift, stellar mass and luminosity affecting the dust mass 
(and hence the gas mass) measurement and distribution, we adopted a stacking technique to derive
the average FIR fluxes in bins of stellar mass and redshift. This method ensures a high level of
completeness, since only a small
fraction of sources are individually detected by Herschel, especially in the longest wavelength bands.

\subsection{Stellar Mass} \label{Ms}
Stellar masses for AGN-hosting galaxies in the sample were derived using the same method as in \cite{Santini12}, where
several consistency tests were also performed. The observed optical photometry was fitted with a combined library of
stellar synthetic templates \citep{Bruzual03} and pure AGN emission templates affected by levels of absorption
\citep{Silva04} associated with different column densities $N_H$. 

In order to break the degeneracies between the different fit parameters, the AGN were classified
into Type I (unabsorbed) and Type II (absorbed) sources on the basis of optical spectroscopic classification
(where available), best-fitting Spectral Energy Distribution (SED) template from \cite{Salvato11}
and intrinsic absorption derived from X-ray data analysis \citep{Lanzuisi13, Mainieri07, Mainieri11,Bauer04}.
Sources for which all these priors are lacking, as well
as X-ray undetected AGN (i.e. Spitzer selected AGN) were considered as Compton-Thin Type II AGN. Moreover, by averaging the two templates of moderate absorption, the four \cite{Silva04} templates were reduced to three: unabsorbed ($N_H<10^{22}\rmn{cm^{-2}}$), Compton-Thin absorbed ($10^{22}<N_H<10^{24}\rmn{cm^{-2}}$) and Compton-Thick absorbed ($N_H>10^{24}\rmn{cm^{-2}}$). Only the most suitable template, based on the above classification, was fitted to each AGN. In particular, the unabsorbed, absorbed Compton Thin and absorbed Compton Thick templates were used for 100, 694 and 7 sources, respectively. 

The fit was performed  through a $\chi^2$ minimization up to $5\,\mu m$ rest-frame, since the \cite{Bruzual03} templates do not include emission from dust reprocessing, assuming an exponentially declining star-formation history and a Salpeter IMF for the stellar component. Each flux was weighted by the inverse of the photometric error and the redshift of each object was fixed during the fit. The stellar mass was derived by the best-fitting stellar component alone.%

 \subsection{Stacking procedure}\label{stack}
 AGN and galaxies are divided into 6 stellar mass bins (from $\rmn{log(M_*/M_\odot)}=9$ to 12) and 4 redshift bins. The number of AGN and galaxies in each bin is reported in Table ~\ref{tab1}. Hereafter, the stellar mass corresponding to each bin is assumed to be the mean stellar mass of the objects in the bin, with uncertainty equal to one standard deviation (typically $\sim 0.14$ dex for both AGN and galaxies). 
 We adopt the stacking procedure implemented by \citeauthor{Santini13},  where details on the stacking procedure
 are given (see also \citealt{Shao10}, \citealt{Rosario12}
 and \citealt{Santini12}). However, in contrast with Santini et al.,
 we do not bin in Star Formation Rate (SFR), since our goal is to investigate if differences
 in terms of gas content between AGNs and normal galaxies may actually
 be at the origin of their claimed SFR differences,
 hence we do not want to include any {\it a priori} selection or binning on the SFR.
 
 Details on the stacking procedure are given in \cite{Santini13}.
 Here we only briefly summarize the basic steps.
 For each Herschel band we excluded areas on the map where the integration time is lower
than half the maximum, to avoid regions of high noise level.
Then, for each $z - M_*$ bin containing at least 10 objects,
we stack on the residual image (in which all the 3$\sigma$ detected
objects were subtracted) of each Herschel band at the position of the undetected sources in the bin, weighting with
the inverse of the square of the error map. Fluxes on the
stacked images were measured through a PSF fitting (for
the PACS bands) or from the value of the central pixel (on
the SPIRE images). Errors on the stacked fluxes were computed through a bootstrap procedure. Finally, the average
flux S in each Herschel band is computed as:

\begin{equation}
 S = \frac{S_{stack}\times N_{stack} + \sum^{N_{det}}_{i=1}S_i}{N_{stack}+N_{det}}
\end{equation}

where $S_{stack}$ is the stacked flux of the $N_{stack}$ undetected
objects in the bin and $S_i$ is the flux of each of the $N_{det}$
detected objects at the $3\sigma$ confidence level.

   \begin{table*}
\caption{Number of AGN and galaxies (between brackets) stacked in each $z-\rmn{M_*}$ bin (see \S~\ref{stack}). The bins that fulfil the conditions in \S~\ref{Md} are in boldface.}\label{tab1}
\begin{tabular}{|r|r|r|r|r|r|r|}
\hline
  \multicolumn{1}{|c|}{ z } &
  \multicolumn{6}{|c|}{{\bf $\rmn{log(M_*/M_\odot)} $ }} \\
        &  9-10   & 10-10.5  & 10.5-11  &  11-11.25  &  11.25-11.5    & 11.5-12      \\

  \hline
  0.0-0.3      &  \bf{10} ({\bf 3120}) & $<10$ ({\bf 735})    &  \bf{17} ({\bf 496})  &  $<10$ ({\bf 125})    & $<10$  ({\bf 37})     & $<10$ ($<10$)         \\
  0.3-0.6      &  \bf{48} ({\bf 13800}) & \bf{32}({\bf 3881})  &  \bf{65} ({\bf 2795}) &  \bf{42} ({\bf 801})  &  \bf{22} ({\bf 414})  &\bf{16  }({\bf 117})    \\
  0.6-0.8      &  29 ({\bf 10621})      & \bf{24} ({\bf 5065}) &  \bf{61} ({\bf 3071}) &  43 ({\bf 1018})       & \bf{42} ({\bf 504})   & 16 ({\bf 199})    \\
  0.8-1.0      &  17 ({\bf 6767})      & \bf{19} ({\bf 6209}) &  \bf{72} ({\bf 4393}) &  \bf{70} ({\bf 1372}) &  \bf{53} ({\bf 754})  & \bf{30} ({\bf 295})   \\

  \hline

\end{tabular}

\end{table*}

 \subsection{Dust Mass}\label{Md}
FIR emission in star forming galaxies is mainly due to galactic cold dust, heated by young stars. The dust mass can be
inferred by fitting dust emission models to the FIR fluxes varying the temperature
distribution and normalization of the far-IR SED.
We considered only $z-\rmn{M_*}$ bins in which the stacked flux has a $3\sigma$ significance in at least three Herschel
bands, at least one of which between the $350$ and $500\,\rmn{\mu m}$ bands.
This requirement ensures a good sampling of the dust emission peak and, hence, reliable estimates of the dust mass.

We performed a $\chi^2$ minimization to the analytical SED templates by \cite{Draine07} (which
span a broad range of dust temperature distributions) and, in each $z-\rmn{M_*}$ bin, the
best-fitting dust mass ($M_d$) was derived from the normalization of the template corresponding to the minimum $\chi^2$. 
Further details on the method are given in \citeauthor{Santini13}.
Errors were estimated from the range of $M_d$ covered by all the templates within $\Delta\chi^2 = 1$ from the value of the
best fitting template. 
Regarding the AGN sample, since emission from AGN-heated dust could contribute to the FIR fluxes,
especially in the PACS bands, we added the \cite{Silva04} AGN SEDs during the \cite{Draine07} templates fit, leaving their normalization free to vary. However, we
shall mention that the results on the dust masses are
essentially unchanged even by neglecting the AGN contribution in the SED fitting; this is not surprising, not only because
AGN heating contribute mostly at mid-IR wavelengths (which are not relevant for the dust mass derivation), but
also because most of the AGNs in our sample do not have extreme luminosities (see \S~\ref{sec2}).

\subsection{Gas mass}\label{Mg}
The total gas mass ($M_{gas}$, which incorporates both the molecular and atomic phases) can be inferred from the dust mass through a dust-to-gas ratio (DGR; e.g. \citealt{Eales10,Magdis12}): $M_{gas}=M_d/\rmn{DGR}$. In order to derive the DGR, we assumed that a fixed fraction of metals is incorporated into dust grains (e.g. \citealt{Draine07b} and references therein; \citealt{Leroy11,Smith12,Corbelli12,Sandstrom12}) and that the DGR scales linearly with the gas metallicity, traced by the oxygen abundance \citep{Draine07b}:
\begin{equation}
 \rmn{DGR} = 0.01\cdot 10^{Z-Z_\odot}.
\end{equation}
This assumption
is observationally found to be a good approximation down to metallicities Z=12+log(O/H)$\sim $8.0 \citep{Remy-Ruyer13}, which is certainly lower than the metallicity range spanned by our sample.
We refer to \citeauthor{Santini13} for a more extended discussion on this method to derive the gas mass from the dust mass.

We derived the mean gas metallicity ($Z$) in each $z-\rmn{M_*}$ bin from the Fundamental Metallicity Relation (FMR) by \cite{Mannucci10,Mannucci11}, who found a
narrow (scatter of $\sim 0.05\,\rmn{dex}$) relation between $M_*$, SFR and $Z$ in local and high-redshift galaxies. 

The average SFR for each $z-\rmn{M_*}$ bin was computed from the total 8--1000$\mu$m luminosity ($L_{IR}$ ), derived by integration of the best-fitting dust emission template used in \S~\ref{Md}:

\begin{equation}
 \rmn{SFR[M_\odot/yr] = 1.8\times10^{-10}\,L_{IR}[L_\odot]}.
\end{equation}

Stellar mass and SFR were converted from a Salpeter to a Chabrier IMF  \citep{Davè08,Santini12_2}, as \cite{Mannucci10,Mannucci11} assumed. Errors on $Z$ were estimated considering the FMR intrinsic scatter and the errors on $M_*$. Errors on SFR were ignored, since they are negligible with respect to the other uncertainties (typically $\Delta\rmn{logSFR}/\rmn{logSFR}\simeq0.04$). Finally, errors on $M_d$ and $Z$ were propagated to derive the uncertainties on $M_{gas}$.
We shall mention that if the FMR is ignored and the metallicities are simply inferred from the mass-metallicity relation \citep{Tremonti04}
and assuming a redshift evolution of this relation \citep{Maiolino08,Troncoso13}, the results remain
unchanged. More generally the bulk of the M$_{gas}$ variations are due to the variations of M$_{dust}$, while variations
and uncertainties in the DGR play a secondary role (i.e. the main results do not change even
if we only consider the dust mass instead of the gas mass, as we will discuss later on).

We also note that the effect of the presence of an AGN on the DGR, if any, would be to decrease its
value with respect to inactive galaxies. Indeed, the hard nuclear radiation is expected to make the
environment less favourable to dust survival. Nonetheless, we assumed the same DGR for AGN hosts as for inactive galaxies. As our results will show, this turns out to be a conservative approach.

\section{Results and discussion}

  \begin{figure*}
\includegraphics[width=80mm,keepaspectratio]{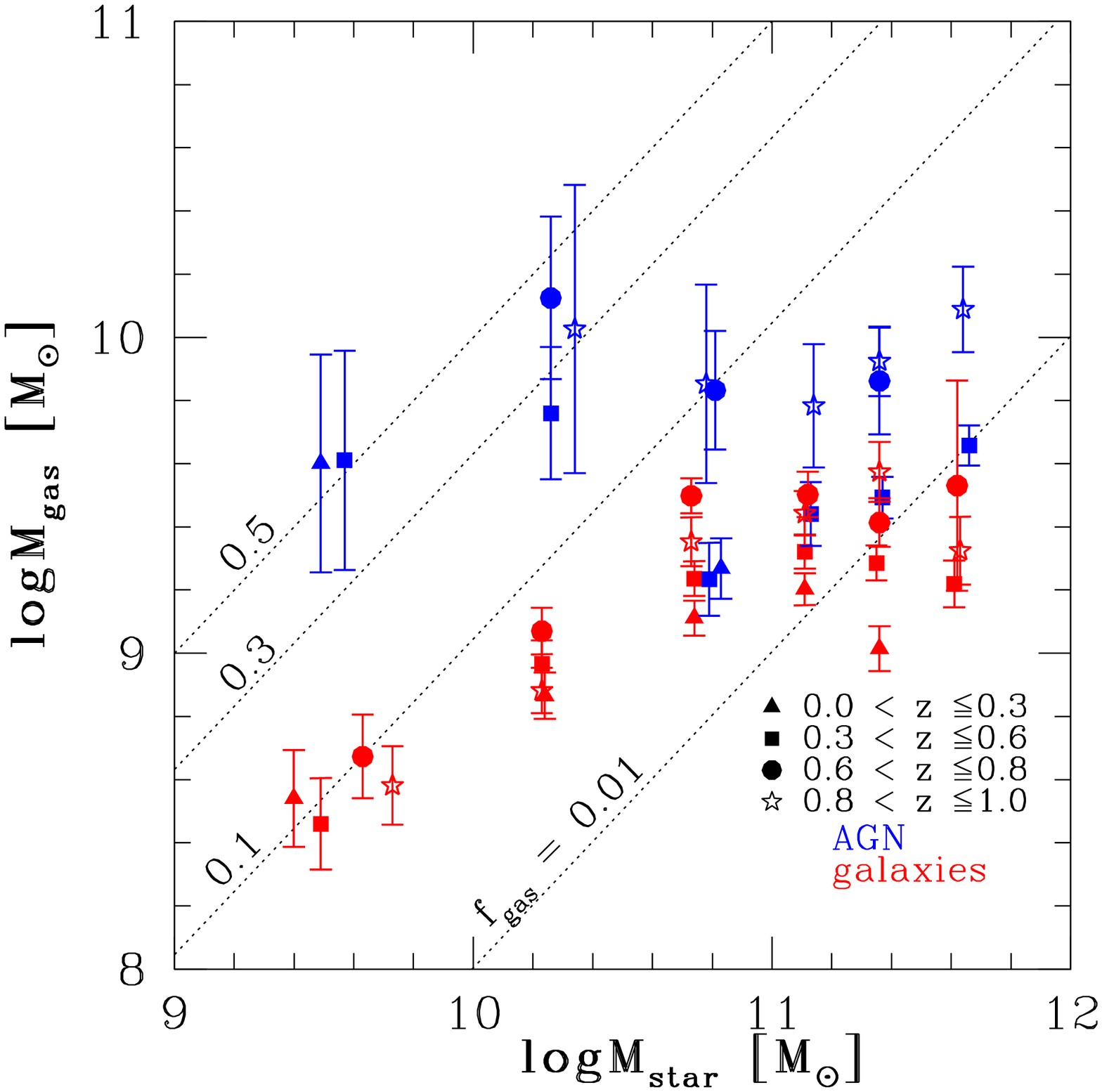}
\includegraphics[width=80mm,keepaspectratio]{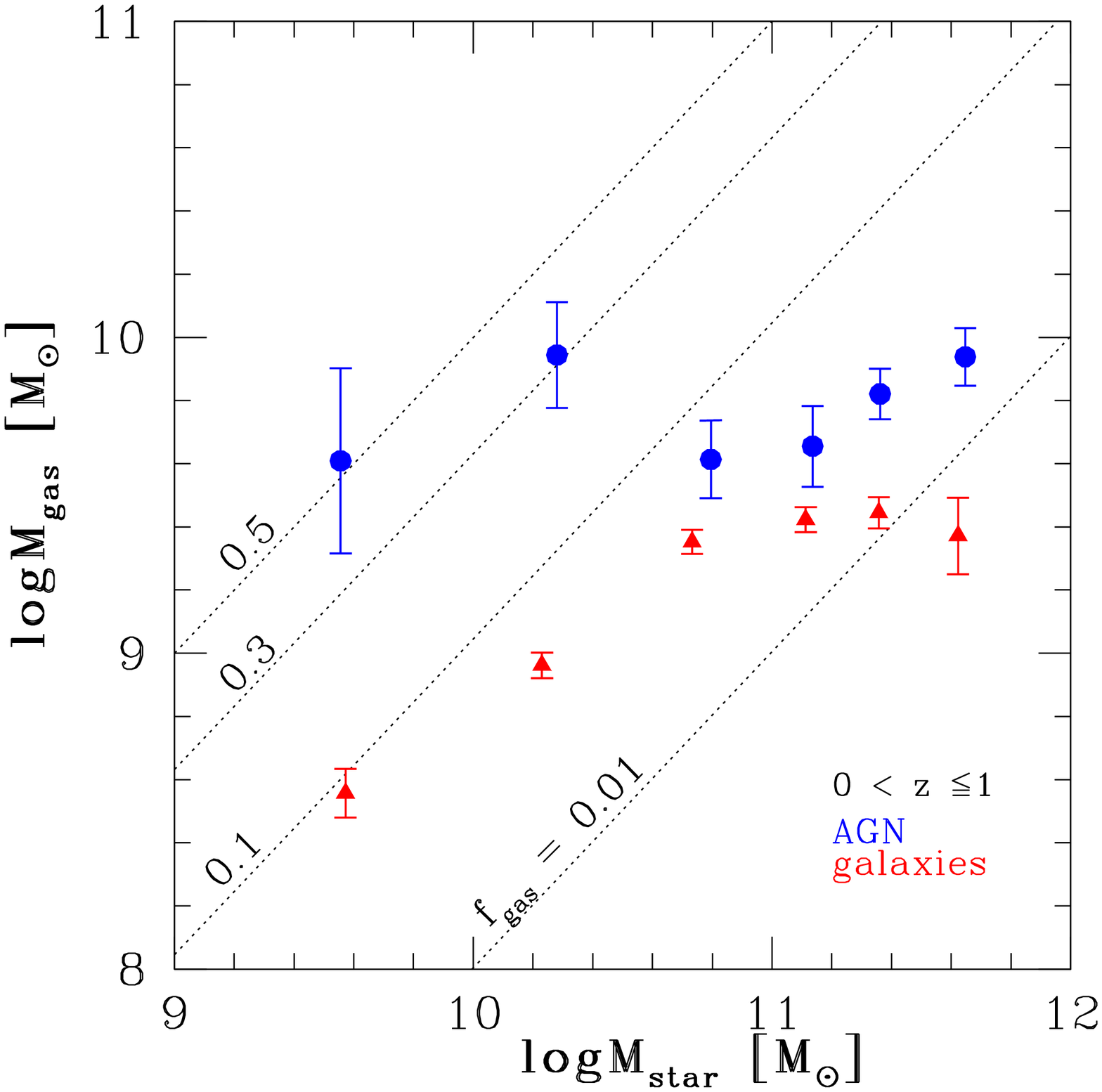}

\caption{Gas mass as a function of the stellar mass for the AGN (blue symbols)
and normal galaxy (red symbols) samples, in different redshift bins (identified
by different symbols; left panel) and averaged over the redshift bins (right
panel). Loci of constant gas fraction, for different values, are shown with dotted lines. }
\label{fig1}
 \end{figure*}

Fig.~\ref{fig1} (left panel) shows the average gas mass as a function of the stellar mass of the AGN and galaxy samples for each
$z-\rmn{M_*}$ bin which fulfils the requirements described in \S~\ref{Md}. The gas mass of galaxies increases with stellar mass, while the gas fraction, defined as $f_{gas}=M_{gas}/(M_{gas}+M_*)$, increases towards lower stellar masses, as already reported by \citeauthor{Santini13}. 

The most interesting result is that
the AGN hosts gas masses are systematically higher than in the galaxy sample, in nearly all bins of stellar mass and
redshift.
Since this result does not show any evident dependence on redshift, in Fig.~\ref{fig1} (right panel) we show the gas mass in the different stellar mass bins, averaged over the redshift bins using weights corresponding to the number of objects in each bin. Largely consistent results are found by computing the unweighted averages.
The same qualitative relations hold if one considers the directly-measured dust mass instead of the more physically relevant gas mass. Indeed, as also reported in \S~\ref{Mg}, gas mass derivation is strongly dominated by the dust mass, with variations on the DGR being second order effects.

\begin{figure}
\includegraphics[width=80mm,keepaspectratio]{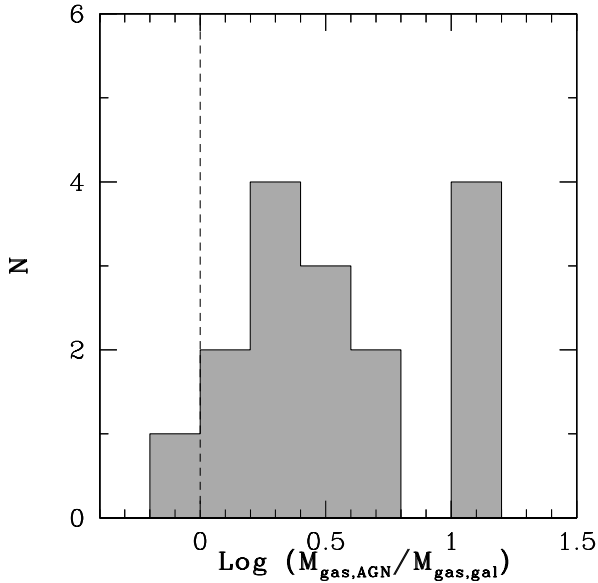}

\caption{Distribution of the ratio between the gas mass in AGN hosts and
in normal
galaxies, inferred within the same $z-\rmn{M_*}$ bins.}
\label{fig2}
 \end{figure}

\begin{figure}
\includegraphics[width=80mm,keepaspectratio]{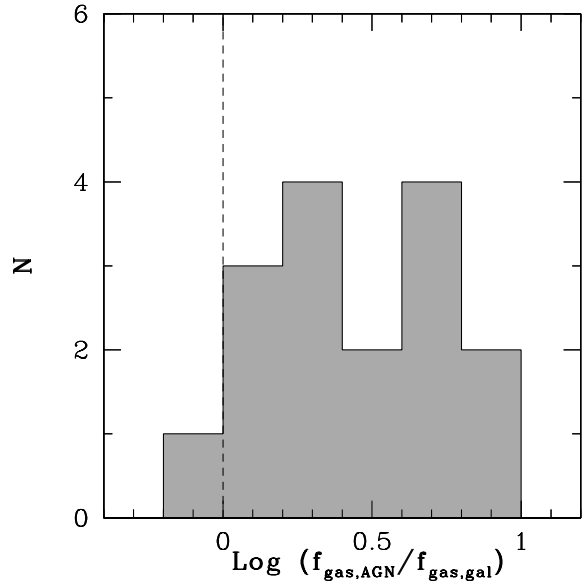}

\caption{Distribution of the ratio between the gas fraction in AGN hosts and
in normal galaxies, inferred within the same $z-\rmn{M_*}$ bins.}
\label{fig3}
 \end{figure}

Fig.~\ref{fig2} shows the distribution of $\rm M_{gas,AGN}/M_{gas,gal}$, for all redshift and
stellar mass bins
for which this comparison can be made. Clearly, in nearly all bins (more specifically, in 94\% of them),
$\rm M_{gas,AGN}/M_{gas,gal} >1$. In 44\% of the bins the gas mass in AGN hosts is higher by more
than a factor of three with respect to normal galaxies. The same result is shown in Fig.~\ref{fig3} in terms of gas fractions,
$\rm f_{gas,AGN}/f_{gas,gal}$.

To better quantify the significance of the result, Table ~\ref{tab2} gives the mean $\rm
M_{gas,AGN}/M_{gas,gal}$ ratio, obtained by averaging the results in three different stellar mass ranges.
The strongest difference is clearly at $\rm \log{(M_{star}/M_{\odot})} < 10.5$, where the gas mass
of AGN hosts is
on average an order of magnitude higher than in normal galaxies, with a significance of $6.5\sigma$. However, the
difference between AGN and normal galaxies in terms of gas masses is significant at $\sim 3-7
\sigma$ also at $\rm \log{(M_{star}/M_{odot})} > 10.5$.

We checked that no significant change in the results are obtained considering different mass or redshift bins. The low stellar-mass bins can be affected by incompleteness, also because the K-band selection preferentially rejects low mass objects. However, the same level of completeness is expected to characterize the AGN and galaxy samples, as the K-band is dominated by the stellar component.
The main conclusion holds, besides the loss of statistics, even repeating the analysis separately on X-ray detected and undetected AGN, as well as applying different cuts in $L_X$.

   \begin{table}
\caption{Mean $\rmn{\log{( M_{gas,AGN}/M_{gas,gal})}}$ (computed in $z-\rmn{M_*}$ bins) for three different mass ranges.}\label{tab2}
\begin{tabular}{|r|r|r|r|}
\hline

  \multicolumn{1}{|c|}{{\bf{$\rmn{\log{(\frac{M_*}{M_\odot}})}$}}} &
  \multicolumn{1}{|c|}{  9-10.5 } &
 \multicolumn{1}{|c|}{ 10.5-11.25 } &
 \multicolumn{1}{|c|}{{    11.25-12  }} \\

  \hline
  \bf $\rm \langle \log{( \frac{M_{gas,AGN}}{M_{gas,gal}})}\rangle$       &  $1.04\pm0.16$    &
  $0.24\pm0.08$ &  $0.44\pm0.06$    \\
  \hline
\end{tabular}
\end{table}

 Since gas accretion onto Super Massive Black Holes (SMBH) is the process at the origin of 
 nuclear activity, whatever the mechanisms that drives the gas into the central regions are, it is not surprising that
 AGNs are preferentially hosted by gas rich galaxies \citep[see also][]{Silverman09}. Indeed, the conditions for gas
 accretion are statistically easier to be fulfilled in presence of a larger gas content. Beyond
 these simple
 statistical arguments, models have been proposed that ascribe AGN secular fuelling
 to disk instabilities \citep{Bournaud11}, which are stronger in gas rich disks. Our results support this scenario.

As discussed above, the gas content is also the fundamental ingredient driving star formation in galaxies,
through the SK relation.
Several works have found a relation between strong nuclear activity and enhanced SFR with respect to inactive
galaxies \citep[e.g][]{Lutz08, Lutz10, Shao10, Rosario12}. \cite{Rosario13_2} and \cite{Santini12} found that the
Herschel detection fraction for AGN is higher than for galaxies and concluded that AGN are more likely to be hosted
by star forming galaxies.

Given the dependency of the SFR on the gas content and the
result obtained by us, the enhanced star formation in AGN galaxies appears
to be primarily the result of a larger (on average) gas content, with respect to the 
bulk of the galaxy population (star forming and quiescent) at similar stellar masses, as already suggested by \cite{Rosario12, Rosario13_2} and \cite{Santini12}.
However, the differences in terms of SFR are less
clear with respect to the differences found by us in terms of gas content, probably because of the additional spread
introduced by the SK relation, and the contribution of triggering mechanisms (e.g. galaxy
interactions),
which may affect the star formation efficiency per unit gas mass. It is beyond the scope of
this paper to further discuss the level and nature of enhanced SFR in AGN hosts.

The main important result of our work is that AGN host galaxies are characterized by much larger amounts of gas,
strongly suggesting that generally AGN activity in galaxies is simply fostered by a larger content
of gas, without invoking specific triggering mechanisms.

\section{Conclusions}
 Making use of the wide multiwavelength data available in the COSMOS, GOODS-S and GOODS-N fields, we
 selected a sample of AGN and galaxies at $z<1$. A stacking procedure on the Herschel maps was
 implemented to derive the average Herschel fluxes in bins of stellar mass and redshift. The stacked
 FIR fluxes were then used to derive the average dust mass and, under reasonable assumptions, the
 average gas mass of AGN hosts and normal galaxies. Finally we compared the average gas mass of AGN hosts and inactive galaxies in the same $z-\rmn{M_*}$ bins.
 
 We find that, at a given stellar mass and redshift, AGNs are hosted in galaxies
 much more gas-rich than inactive ones.
The difference is strongest at low stellar masses,  $\rmn{log(M_*/M_\odot)}<10.5$, where the
gas mass in AGN hosts is on average ten times higher than in normal galaxies (a result significant at $6.5\sigma$).
Significantly higher gas masses, relative to the normal galaxy population,
are however also observed in AGN hosts with stellar masses higher
than $\rmn{log(M_*/M_\odot)}>10.5$. Taken altogether, in nearly all
stellar mass and redshift bins AGN host galaxies have higher gas content than normal galaxies;
in almost half of the sample the gas fraction of AGN host galaxies is more than three times higher than
in normal galaxies.

Our result strongly suggests that the likelihood of having an AGN in a galaxy is primarily given by the
amount of gas in the host galaxy, while dynamical triggering processes (bars, galaxy mergers
and interactions) likely play a secondary role, at least
in the luminosity range probed by us. This result can be interpreted in simple statistics terms that it is more
likely that a gas cloud falls into the potential of the supermassive black hole if there are overall
more gas clouds in
the host galaxy. More elaborated models, in which secular fuelling of AGNs is caused by disk instabilities, which
are stronger in more gas rich disks, are also supported by our results.

\section*{Acknowledgments}
We acknowledge support from the Italian Space
Agency under the ASI-INAF contract I/009/10/0 and from INAF under the con-
tract PRIN-INAF-2012. FV thanks B. Luo for kindly providing Chandra ID's of MUSIC counterparts in GOODS-S and I. Delvecchio for useful discussion on SED fitting procedure.
This paper uses data from \textit{Herschel}'s photometers PACS and SPIRE. PACS has been developed by a consortium of institutes led by MPE (Germany) and including UVIE (Austria); KU Leuven, CSL, IMEC (Belgium); CEA, LAM (France); MPIA (Germany); INAF-IFSI/OAA/OAP/OAT, LENS, SISSA (Italy); IAC (Spain). This development has been supported by the funding agencies BMVIT (Austria), ESA-PRODEX (Belgium), CEA/CNES (France), DLR (Germany), ASI/INAF (Italy), and CICYT/MCYT (Spain). SPIRE has been developed by a consortium of institutes led by Cardiff University (UK) and including Univ. Lethbridge (Canada); NAOC (China); CEA, LAM (France); IFSI, Univ. Padua (Italy); IAC (Spain); Stockholm Observatory (Sweden); Imperial College London, RAL, UCL-MSSL, UKATC, Univ. Sussex (UK); and Caltech, JPL, NHSC, Univ. Colorado (USA). This development has been supported by national funding agencies: CSA (Canada); NAOC (China); CEA, CNES, CNRS (France); ASI (Italy); MCINN (Spain); SNSB (Sweden); STFC, UKSA (UK); and NASA (USA).

\bsp

\label{lastpage}

\end{document}